\documentclass[12pt]{article}

\usepackage{graphicx}
\usepackage{amsmath}
\usepackage{amsfonts}
\usepackage{amssymb}

\def\la{\mathrel{\mathpalette\fun <}}

\def\fun#1#2{\lower3.6pt\vbox{\baselineskip0pt\lineskip.9pt
\ialign{$\mathsurround=0pt#1\hfil##\hfil$\crcr#2\crcr\sim\crcr}}}

\title{Tevatron constraints on the Higgs boson mass in the
fourth-generation fermion models revisited }
\author{A.N. Rozanov\thanks{CPPM
(IN2P3-CNRS-Universite de Mediterranee), Marseille, France}, \\
M.I. Vysotsky\thanks{ITEP, Moscow, Russia}}
\date{}

\begin{document}
\maketitle
\begin{abstract}
Recent Tevatron exclusion interval of the masses of Higgs boson
considerably reduces in case of the light quasistable fourth generation
neutral lepton.

\end{abstract}

If the fourth sequential quark-lepton generation does exist then
the cross section of Higgs boson production at hadron colliders is
considerably enhanced in comparison with that in Standard Model
(SM) \cite{1}. This result was used in a recent Tevatron paper
according to which a standard-model-like Higgs boson in the mass
interval

\begin{equation}
131 \; {\rm GeV} \; < m_H < 204 \; {\rm GeV} \;\;  \label{1}
\end{equation}
is excluded at the 95\% Confidence Level in
the model with the fourth generation \cite{2}. The statement about
exclusion follows from Fig. 4c of \cite{2}, where an experimental
upper bound on the product $\sigma(gg\to H) \times {\rm Br}(H\to
W^+ W^-)$ is compared with the theoretical prediction for this
product.

The result obtained in \cite{2} strongly depends on the lower mass
bounds on the fourth generation fermions. The point is that the
new decay channel $H\to f\bar f$ opens if a mass of any of these
new particles is less than $m_H/2$. Then ${\rm Br}(H\to W^+ W^-)$
diminishes and the exclusion interval of $m_H$ reduces. Concerning
the fourth generation quarks we know from Tevatron that their
masses are larger than 300 GeV \cite{3}. The mass of the charged
lepton $m_{E}$ is bounded to be above 100 GeV by LEP II, so the
decay $H\to E^+ E^-$ practically does not occur for $m_H$ from
the excluded domain. For the fourth generation neutrino a lower
bound on its mass $m_{N} > 80$ GeV obtained at LEP II \cite{4}
is used in \cite{2}. In \cite{2} two scenarios are considered:
$m_{N} = 80$ GeV (low mass scenario) and $m_{N} \gg 80$
GeV (high mass scenario). The above mentioned exclusion interval
of $m_H$ refers to low mass scenario; for high mass scenario an
exclusion interval of $m_H$ stretches till $m_H = 208$ GeV.

The aim of the present note is to stress that a lower bound
$m_{N} > 80$ GeV \cite{4} is applicable only to the case
when the mixing angle of the
fourth generation neutral lepton with at least one neutral lepton
from three light generations is larger than $3\cdot
10^{-6}$.
In this case $N$ decays to charged leptons from the first three generations inside
L3 detector.
 For smaller mixing angles (quasistable $N$) the mass
of $N$ is bounded only from the analysis of $Z$ boson decays,
$m_{N} > 46.7$ GeV \cite{5}.\footnote{Since $N$ is at
least $10^{11}$ times heavier than the heaviest of three SM
neutrinos, a values of the lepton mixing angles $\theta_{i4} \approx
\sqrt{m_{\nu i}/m_{N}} < 3\cdot 10^{-6}$ look quite natural.}
If the decay of Higgs boson to a pair of heavy neutral leptons is
kinematically allowed, then it dominates \cite{6}. In \cite{7} we
study how Standard Model Higgs boson branching ratios is
changing in the presence of
light $N$.

In Fig. 1 we compare the branching ratios of Higgs to WW
calculated with modified HDECAY code \cite{66} for $m_{N}= 80$
GeV, $m_E=100$ GeV, $m_U=450$ GeV, $m_D=400$ GeV (black curve)
with the branchings used in \cite{2} (red curve). The agreement
between two calculations is very good. In Fig. 2 the same
branching ratios for $m_{N} = 46.7$ GeV are shown.

In the Table we present the branching ratios of $H\to W^+ W^-$
decays for $m_N = 80$ GeV and for $m_N = 46.7$ GeV for $m_H$ from
110 to 300 GeV.

\newpage

\begin{center}

{\bf Table}

\bigskip

\begin{tabular}{|c|c|c|}
\hline && \\
$m_H$ (GeV) & Br ($H\to W^+ W^-$) & Br ($H\to W^+ W^-$) \\
& $m_N = 80$ GeV & $m_N = 46.7$ GeV \\
\hline
110 & 0.03 & 0.005 \\
120 & 0.08 & 0.01 \\
130 & 0.19 & 0.02 \\
140 & 0.35 & 0.04 \\
150 & 0.55 & 0.10 \\
160 & 0.85 & 0.37 \\
170 & 0.88 & 0.68 \\
180 & 0.83 & 0.73 \\
190 & 0.69 & 0.67 \\
200 & 0.65 & 0.65 \\
210 & 0.62 & 0.63 \\
220 & 0.60 & 0.62 \\
230 & 0.59 & 0.61 \\
240 & 0.58 & 0.61 \\
250 & 0.58 & 0.60 \\
260 & 0.58 & 0.60 \\
270 & 0.57 & 0.60 \\
280 & 0.58 & 0.60 \\
290 & 0.58 & 0.60 \\
300 & 0.58 & 0.60 \\
\hline
\end{tabular}

\bigskip

{\it Branching ratios of $H\to W^+ W^-$ decays for two values of
$m_N$.}

\end{center}

\bigskip

From the Table we see, that branching ratio of the decay $H\to W^+
W^-$ considerably diminishes for $m_H < 160$ GeV. Taking this
effect into account from Figures 4(c-d)  and Tables I-II of
\cite{2}  we obtain the following model independent exclusion
interval:
\begin{equation}
155\; {\rm GeV} \; < m_H < 204 \; {\rm GeV \; excluded \; at \;
95\% \; C.L.} \;\;  . \label{2}
\end{equation}

Our second comment refers to the case of $m_{N} > 80$ GeV.
Fourth generation change considerably the constrains on $m_H$
from electroweak precision data. In particular, one can
choose the values of the fourth generation masses so,
that heavy Higgs is allowed. Only
an upper bound
$m_H \la 1$ TeV from perturbative unitarity \cite{8} remains.


\begin{center}
\bigskip
\includegraphics[width=.8\textwidth]{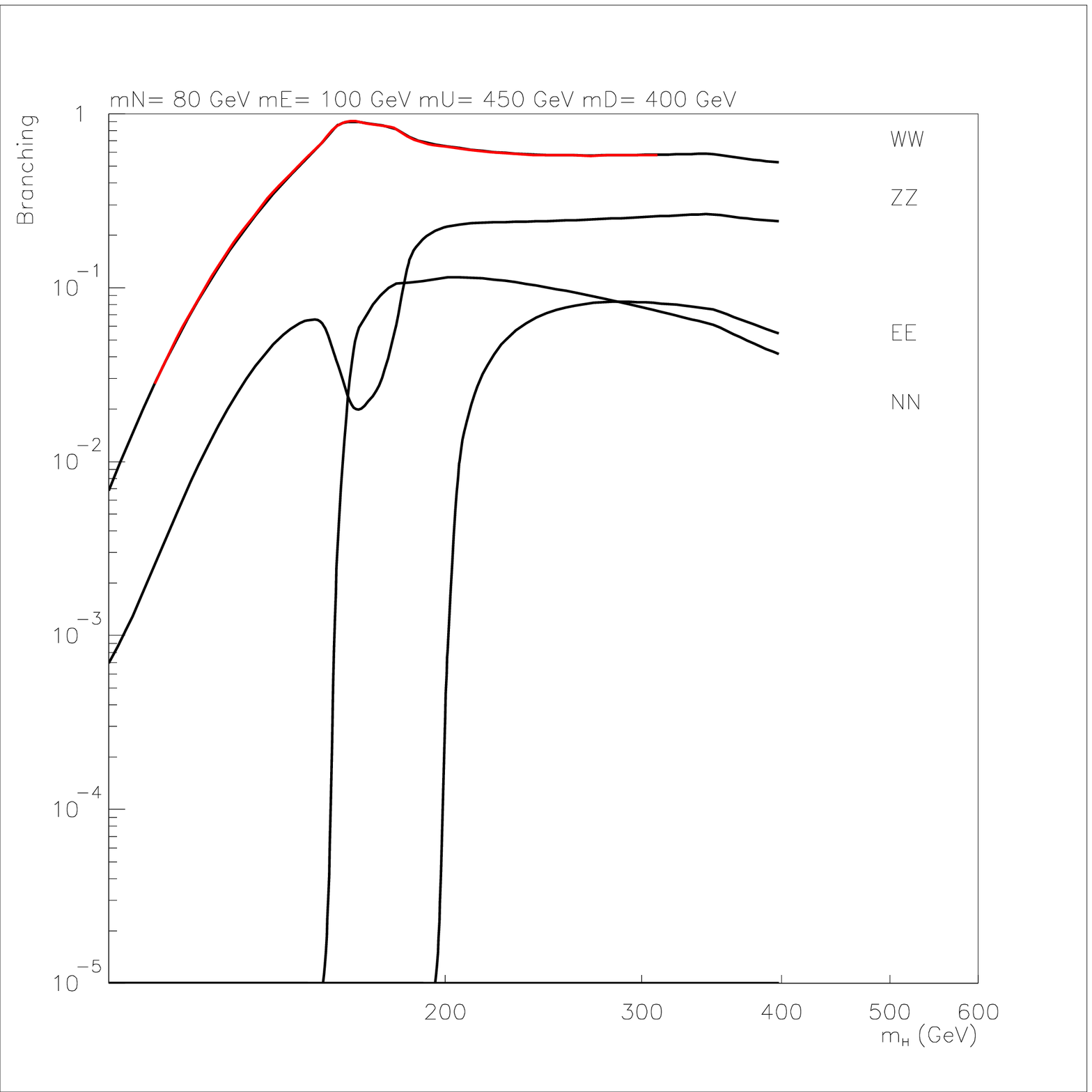}

Fig. 1. {\em Branching ratios of Higgs boson decays in case
of fourth generation with $m_{N} = 80$ GeV. Red line
corresponds to the branching ratios  from the last column of Table I
of the CDF-D0 paper \cite{2}. (The values
$m_E=100$ GeV, $m_U=450$
GeV, $m_D=400$ GeV are used).}

\end{center}

 In
\cite{9} we study the value of $m_H$ (where
minimum of $\chi^2$ of the electroweak data fit occurs)
as a function of the mass of the neutral
lepton $N$. According to Fig. 5 from \cite{9} for the case of one
extra generation and the fourth lepton heavier than 80 GeV, Higgs
boson mass less than 240 GeV corresponds to the $\chi^2$ minimum.
It would mean that a considerable part of the allowed interval of
$m_H$ is depreciated by the bound (\ref{1}) valid for $m_{N} > 80$
GeV.
However, in the analysis of paper \cite{9} we neglect a possible
CKM type mixing of the fourth generation quarks with the quarks of
three ``light'' generations. This mixing was taken into account in
the recent paper \cite{10} with the following result: for $m_N =
101$ GeV and sine of quark mixing angle $s_{34} \sim 0.1 \div 0.2$
the value
of $m_H$ up to 600 GeV is allowed.

\begin{center}
\bigskip
\includegraphics[width=.8\textwidth]{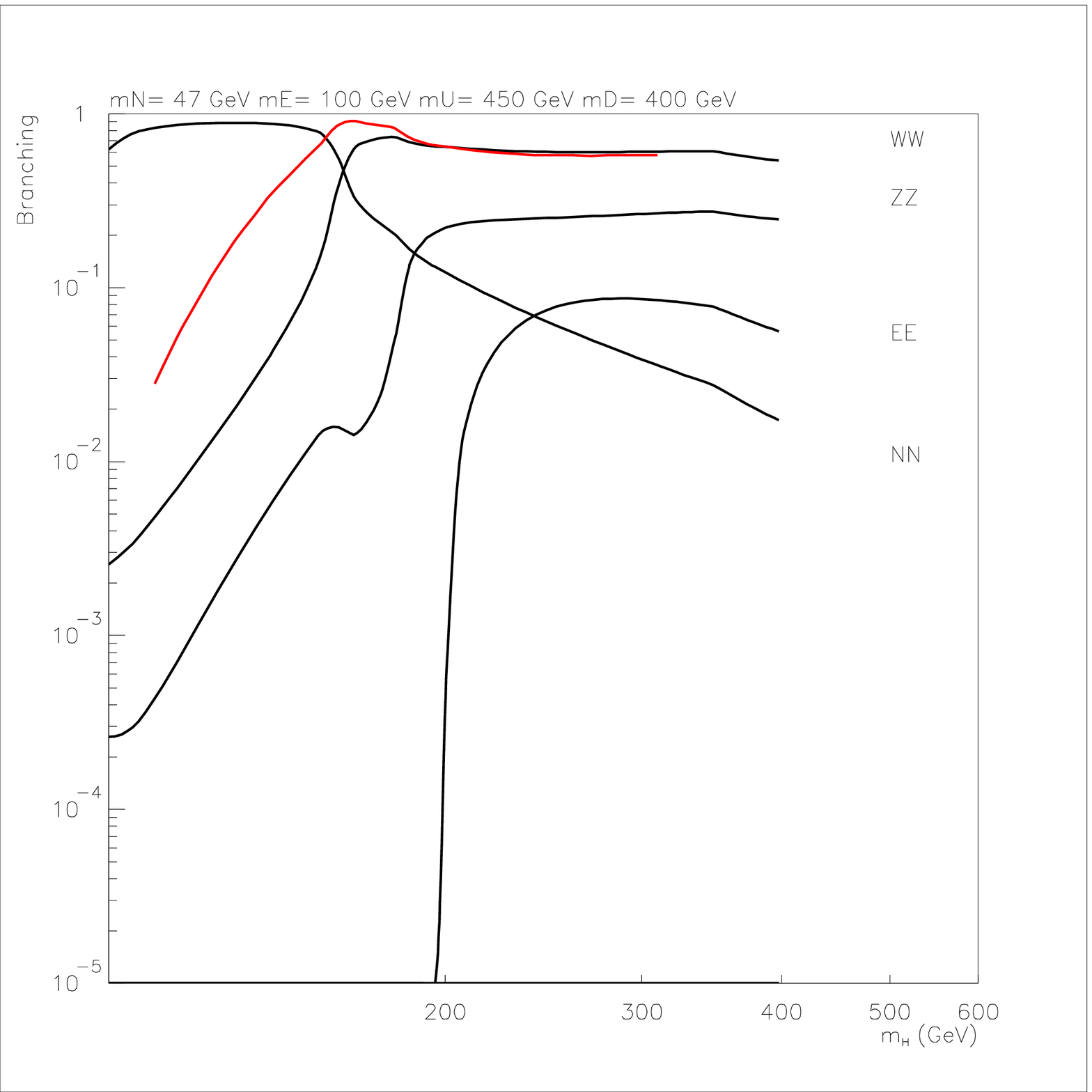}

Fig. 2. {\em Branching ratios of Higgs boson decays in case
of fourth generation with $m_{N} = 46.7$ GeV.
Red line demonstrates growth of the branching ratio of Higgs
decay into $WW$ for $m_{N} = 80$ GeV. (The values
$m_E=100$ GeV, $m_U=450$
GeV, $m_D=400$ GeV are used)}.

\end{center}

 At the absence of mixing
in accordance with our results \cite{9} Tevatron bound (\ref{1})
almost excludes the existence of the fourth generation with heavy $N$.
However the conclusion of \cite{10} that zero CKM mixing $s_{34}$
is excluded is not valid for the interval of
heavy neutrino masses $m_{N} = 46.7 - 70.0$ GeV.

In a very interesting recent paper \cite{11} the fourth generation
with extremely
small mixing with lighter three generations is considered.
The main issue of \cite{11} is the preservation of
baryon and lepton asymmetries against sphaleron erasure in this model.
The fact that the exclusion interval of the higgs boson masses
(\ref{1}) diminishes to (\ref{2}) in case of the quasistable $N$
enlarge the allowed parameter space which could be used in \cite{11}.
The bounds from the EW precision data are discussed for
the case of light $N$ in the $STU$ formalism in \cite{11}.
In \cite{9} we specially
address an issue of inaplicability of $STU$ formalism in the case of light
$N$. The use of the proper parameters ($V_i$ or $S',T',U'$) would
change the allowed domain in the $m_U - m_D$ plot (Fig 1) of  \cite{11}.

In summary, we demonstrated that model independent exclusion interval of
the values of Higgs boson masses from Tevatron direct searches
in case of fourth generation is reduced to $155$ GeV $< m_H < 204$ GeV, by
allowing  small heavy neutrino masses $m_{N} = 45.7-80.0$ GeV.

M.V. is partially supported by the grant N-Sh 4172.2010.2 and the contract
02.740.11.5158 of the Ministry of Education and Science of the RF.

\end{document}